\documentclass{article}\usepackage{ciss2005}

\usepackage{graphicx, comment}
\usepackage{tabularx}
\usepackage{booktabs}
\usepackage{multirow}
\usepackage{amsmath, amsfonts}

\def\BibTeX{{\rm B\kern-.05em{\sc i\kern-.025em b}\kern-.08em
    T\kern-.1667em\lower.7ex\hbox{E}\kern-.125emX}}

\newtheorem{theorem}{Theorem}

\newtheorem{corollary}{Corollary}

\newenvironment{thmproof}[1]
{\noindent\hspace{2em}{\it #1 }}
{\hspace*{\fill}~\QED\par\endtrivlist\unskip}

\def\GF{\mathop{\sf GF}\limits}
\def\bigorder{{\mathcal{O}}}
\def\RR{{\mathbb{R}}}
\def\NN{{\mathbb{N}}}
\def\PP{{\mathsf{P}}}
\def\EE{{\mathsf{E}}}
\def\const{{\mathsf{const}}}

\def \Theorem#1{{\bf Theorem~#1}}

\def \Corollary#1{{\bf Corollary~#1}}

\def \Figure#1{{\rm Fig.~#1}}
\def \Figures#1{{\rm Figs.~#1}}
\def \Table#1{{\rm Table~#1}}

\def \Append#1{{\sc Appendix~#1}}

\def\picchk{\put(0,0){\circle{30}}\put(0,0){\makebox(0,0){\small$+$}}}

\begin{document}

\itwtitle       {On the Typicality of the Linear Code Among the LDPC Coset Code
Ensemble}

\itwauthor      {C.-C.~Wang, S.R.~Kulkarni, and H.V.~Poor\footnotemark[1]}
                {Department of Electrical Engineering \\
         Princeton University \\
         Princeton, New Jersey  08544\\
         e-mail: {\tt \{chihw, kulkarni, poor\}@princeton.edu}}

%

\itwmaketitle

\footnotetext[1]{This research was supported in part by the Army Research
Laboratory under Contract DAAD-19-01-2-0011.}

\begin{itwabstract}
Density evolution (DE) is one of the most powerful analytical
tools for low-density parity-check (LDPC) codes on memoryless
binary-input/symmetric-output channels. The case of non-symmetric
channels is tackled either by the LDPC coset code ensemble (a
channel symmetrizing argument) or by the generalized DE for linear
codes on non-symmetric channels. Existing simulations show that
the bit error rate performances of these two different approaches
are nearly identical. This paper explains this phenomenon by
proving that as the minimum check node degree $d_c$ becomes
sufficiently large, the performance discrepancy of the linear and
the coset LDPC codes is theoretically indistinguishable. This
typicality of linear codes among the LDPC coset code ensemble
provides insight into the concentration theorem of LDPC coset
codes.
\end{itwabstract}


\begin{itwpaper}

\itwsection{Introduction} Low-density parity-check (LDPC) codes
\cite{Gallager63} have found many applications in cellular networks,
magnetic/optical storage devices, and satellite communications, due to their
near-capacity performance and the embedded efficient distributed decoding
algorithms, namely, the belief propagation (BP)
decoder~\cite{McElieceMacKayCheng98}. For binary-input/symmetric-output (BI-SO)
channels, the behavior of the BP decoder and the decodable noise threshold can
be explained and predicted by the density evolution (DE) method, which traces
the evolved distribution on the log-likelihood ratio (LLR) message used in
BP~\cite{RichardsonUrbanke01a}. Additional references on LDPC codes can be
found in~\cite{RichardsonShokrollahiUrbanke01}.

Although the classical DE does not apply to binary-input/non-symmetric-output
(BI-NSO) channels, in practice, LDPC codes are applicable to BI-NSO channels as
well and near capacity performance is reported \cite{WangKulkarniPoor03}.
Rigorous analyses of BI-NSO channels are addressed either by the coset code
argument (namely, a channel-symmetrizing
argument)~\cite{KavcicMaMitzanmacher03,HouSiegelMilisteinPfister03} or by the
generalized DE for linear codes on BI-NSO channels~\cite{WangKulkarniPoor03}.

A coset code consists of all sequences $\mathbf{x}$ of length $n$
satisfying
\begin{eqnarray}
{\mathbf{Hx}}=\mathbf{s},\label{def:coset-codes}
\end{eqnarray}
for some fixed, coset-defining syndrome $\mathbf{s}$, where $\mathbf{H}$ is a
fixed parity-check matrix of dimension $(n(1-R))\times n$ and $R$ is the rate
of this coset code. When $\mathbf{s}=\mathbf{0}$, (\ref{def:coset-codes})
corresponds to a linear code. It has been shown in
\cite{KavcicMaMitzanmacher03} that for sufficiently large $n$, almost all
${\mathbf{s}}\in \{0,1\}^{n(1-R)}$ and almost all $\mathbf{H}$ drawn from the
equiprobable bipartite graph ensemble, the {\it codeword-averaged} performance
can be predicted by the coset-code-based DE within arbitrary precision. If one
further assumes that there is a common independent, uniformly distributed bit
sequence accessible to both the transmitter and the receiver, then a
coset-code-averaged (syndrome-$\mathbf{s}$-averaged) scheme can be obtained as
in~\Figure{\ref{fig:coset-code-arg}(a)}. This coset-code-averaged scheme
in~\Figure{\ref{fig:coset-code-arg}(a)} is equivalent to a linear LDPC code on
the symmetrized channel as demonstrated
in~\Figure{\ref{fig:coset-code-arg}(b)}, of which the error-probability is
codeword-independent. On the other hand, the generalized DE
in~\cite{WangKulkarniPoor03} analyzes the {\it codeword-averaged} performance
when linear codes plus BI-NSO channels are considered as
in~\Figure{\ref{fig:coset-code-arg}(c)}. It is shown
in~\cite{WangKulkarniPoor03} that the necessary and sufficient stability
conditions in both schemes (\Figures{\ref{fig:coset-code-arg}(b) {\rm
and}~\ref{fig:coset-code-arg}(c)}) are identical. Monte Carlo simulations based
on finite-length codes ($n=10^4$) \cite{HouSiegelMilisteinPfister03} further
show that the codeword-averaged performance in
\Figure{\ref{fig:coset-code-arg}(c)} is nearly identical\footnote{That is, it
is within the precision of the Monte Carlo simulation.} to the
performance\footnote{In \Figure{\ref{fig:coset-code-arg}(b)}, the error
probability is codeword-independent so that the all-zero codeword can be
assumed, which drastically simplifies the computation of taking the average
over the entire codebook. } of \Figure{\ref{fig:coset-code-arg}(b)} when the
same encoder/decoder pair is used. The above two facts suggest close a
relationship between linear codes and the coset code ensemble.
\begin{figure*}
\centering \setlength{\unitlength}{.1mm}
\begin{tabular}{c}

\begin{picture}(1000,270)(0,-20)

 \put(-50,50){\vector(1,0){50}}
\put(0,0){\framebox(200,100){\tiny lin.\ LDPC ENC}}
\put(325,50){\picchk}
%
\put(255,120){\framebox(140,50){\tiny Rand.\ Bits}}
\put(325,120){\vector(0,-1){55}}
\put(395, 145){\vector(1,0){330}}
\put(725,145){\vector(0,-1){80}}
%
\put(200,50){\vector(1,0){110}}
\put(340,50){\vector(1,0){110}}
\put(450,0){\framebox(200,100){\tiny Non-sym.\ CH.}}
%
%
\put(650,50){\vector(1,0){60}}
%
\put(725,50){\picchk}
\put(740,50){\vector(1,0){60}}
%
\put(1000,50){\vector(1,0){50}}
 \put(800,0){\framebox(200,100){\tiny lin.\ LDPC
DEC}}

 \put(-25,-20){\dashbox{10}(450,220){}} \put(120,210){\tiny LDPC
Coset ENC} \put(685,-20){\dashbox{10}(340,220){}} \put(750,210){\tiny LDPC
Coset DEC}

\end{picture}
\\ (a) Coset Code Ensemble versus Non-symmetric Channels\\
\begin{picture}(1000,270)(0,-20)

 \put(-50,50){\vector(1,0){50}}
\put(0,0){\framebox(200,100){\tiny lin.\ LDPC ENC}}
\put(325,50){\picchk}
%
\put(255,120){\framebox(140,50){\tiny Rand.\ Bits}}
\put(325,120){\vector(0,-1){55}}
\put(395, 145){\vector(1,0){330}}
\put(725,145){\vector(0,-1){80}}
%
\put(200,50){\vector(1,0){110}}
\put(340,50){\vector(1,0){110}}
\put(450,0){\framebox(200,100){\tiny Non-sym.\ CH.}}
%
%
\put(650,50){\vector(1,0){60}}
%
\put(725,50){\picchk}
\put(740,50){\vector(1,0){60}}
%
\put(1000,50){\vector(1,0){50}}
 \put(800,0){\framebox(200,100){\tiny lin.\ LDPC
DEC}}

\put(225,-20){\dashbox{10}(540,220){}} \put(450,210){\tiny
Symmetric Channel}

\end{picture}
\\(b) Linear Code Ensemble versus Symmetrized Channels\\
\begin{picture}(1000,150)

\put(-50,50){\vector(1,0){50}}
\put(0,0){\framebox(200,100){\tiny lin.\ LDPC ENC}}
%
\put(200,50){\vector(1,0){250}}
\put(450,0){\framebox(200,100){\tiny Non-sym.\ CH.}}
%
\put(650,50){\vector(1,0){150}}
\put(1000,50){\vector(1,0){50}}
 \put(800,0){\framebox(200,100){\tiny lin.\ LDPC
DEC}}
\put(-50,50){\vector(1,0){50}}
\put(0,0){\framebox(200,100){\tiny lin.\ LDPC ENC}}

\end{picture}\\
(c) Linear Code Ensemble versus Non-symmetric Channels
\end{tabular}
\itwfigurecaption{Comparison of the approaches based on the coset code ensemble
and on codeword averaging.\label{fig:coset-code-arg}}
\end{figure*}

This paper addresses this phenomenon by proving that for sufficiently large
minimum check node degree $d_c$, the asymptotic performance (and behavior) of
linear LDPC codes is theoretically indistinguishable from that of the LDPC
coset code ensemble. In practice, the convergence rate of the thresholds of
these two schemes  is very fast (with respect to $d_c$). For moderate $d_c\geq
6$, the discrepancy of the asymptotic thresholds\footnote{The asymptotic
thresholds (with respect to $n$) are obtained by DE/generalized DE rather than
Monte Carlo simulation.} for the linear codes and the coset code ensemble
(\Figures{\ref{fig:coset-code-arg}(b) {\rm and}~\ref{fig:coset-code-arg}(c)})
is within $0.05\%$. Our result shows the typicality of linear codes among the
coset code ensemble. Besides its theoretical importance, we are then on solid
ground when simulating the codeword-averaged linear code performance by
assuming the all-zero codeword in the coset code ensemble.

%
%
\itwsection{Formulation\label{sec:formulations}}

\itwsubsection{Code Ensembles} The coset code ensemble is based on
(\ref{def:coset-codes}), where the coset-defining syndrome $\mathbf{s}$ is
uniformly distributed in $\{0,1\}^{n(1-R)}$ and $\mathbf{H}$ is from the
equiprobable bipartite graph ensemble. To be more explicit, the equiprobable
bipartite graph ensemble ${\mathcal C}^n(d_v,d_c)$ is obtained by putting equal
probability on each of the possible configurations of the regular bipartite
graphs with the variable node degree $d_v$ and the check degree $d_c$, and by
the convention that $H_{j,i}$, the $(j,i)$-th entry of the parity check matrix
$\mathbf{H}$, equals one iff there is an {\it odd} number of edges connecting
variable node $i$ and check node $j$. We can also consider irregular code
ensembles ${\mathcal C}^n(\lambda,\rho)$ such that $\lambda$ and $\rho$ denote
the finite order {\it edge degree distribution} polynomials
    \begin{eqnarray}
    \lambda(z)&=&\sum_k \lambda_kz^{k-1}\nonumber\\
    \rho(z)&=&\sum_k \rho_kz^{k-1},\nonumber
    \end{eqnarray}
where $\lambda_k$ or $\rho_k$ is the fraction of edges connecting to a degree
$k$ variable or check node, respectively. Further details on the equiprobable
bipartite graph ensemble can be found in~\cite{RichardsonUrbanke01a}.

    If we hard-wire ${\mathbf{s}}={\mathbf{0}}$ and still let
    $\mathbf{H}$ be drawn from ${\mathcal
C}^n(\lambda,\rho)$, we then obtain the traditional linear LDPC code ensemble.

\itwsubsection{The Classical Density Evolutions} In this paper, we consider
only the BP decoder such that the passed message $m$ corresponds to the LLR
$m=\ln\frac{\PP(y|X=0)}{\PP(y|X=1)}$. The detailed representation of the
variable and check node message maps is as follows.
    \begin{eqnarray}
    m_0&:=&\ln\frac{\PP(y|x=0)}{\PP(y|x=1)}\nonumber\\
    \Psi_v(m_0,m_1,\cdots,m_{d_v-1})&:=&\sum_{j=0}^{d_v-1}m_j\nonumber\\ 
    \Psi_c(m_1,\cdots,m_{d_c-1})&:=&\gamma^{-1}\left(\sum_{i=0}^{d_c-1}\gamma(m_i)\right),\nonumber
    \end{eqnarray}
where $\gamma:\RR\mapsto\GF(2)\times\RR^+$ is such that
\begin{eqnarray}
\gamma(m)&:=&\left(1_{\{m\leq0\}},\ln\coth\left|\frac{m}{2}\right|\right)=(\gamma_1,\gamma_2)\in\GF(2)\times
\RR^+.\nonumber
\end{eqnarray}

 For BI-SO
channels, the probability density of the messages in any symmetric
message passing algorithm is codeword independent, by which we
mean that for different transmitting codewords, the densities of
the messages are of the same shape and differ only in parities.
 Let $P^{(l)}$ denote  the density of the LLR messages
from variable nodes to check nodes during the $l$-th iteration given that the
all-zero codeword is being transmitted. Similarly, $Q^{(l)}$ denotes the
density of the LLR message from check nodes to variable nodes assuming the
all-zero codeword. The classical DE \cite{RichardsonUrbanke01a} derives the
iterative functionals on the evolved densities  as follows.
    \begin{eqnarray}
    P^{(l)}&=&P^{(0)}\otimes \lambda\left(Q^{(l-1)}\right)\nonumber\\
    Q^{(l-1)}&=&\Gamma^{-1}\left(\rho\left(\Gamma\left( P^{(l-1)}\right)\right)\right),\nonumber
    \end{eqnarray}
where ``$\otimes$" denotes the convolution operator and all scalar
multiplications in $\lambda(\cdot)$ and $\rho(\cdot)$ are replaced by
convolutions as well. The operator $\Gamma$ transforms the distribution on
$\RR$ into a distribution on $\GF(2)\times\RR^+$ based on the measurable
function $\gamma(\cdot)$. $\Gamma^{-1}$ represents the corresponding inverse
transform.

\itwsubsection{Generalized DE and the Coset-Code-Based Approach} For BI-NSO
channels, the error-protection capability is codeword dependent and we cannot
assume that the all-zero codeword is transmitted. This difficulty is
circumvented by the codeword-averaged approach in which we trace pairs of
evolved densities, $\left((P^{(l)}(0), P^{(l)}(1)\right)$ and
$\left((Q^{(l)}(0), Q^{(l)}(1)\right)$, where $P^{(l)}(x)$ denotes the
distribution of the LLR message $m=\ln\frac{\PP(y|X=x)}{\PP(y|X=\bar{x})}$ from
the variable node to the check node during the $l$-th iteration, given that the
transmitting bit at the {\it source} variable node is $x$. $Q^{(l)}(x)$ denotes
the distribution of the LLR message $m=\ln\frac{\PP(y|X=x)}{\PP(y|X=\bar{x})}$
from the check node to the variable node during the $l$-th iteration, given
that the transmitting bit at the {\it destination} variable node is $x$. The
generalized DE for linear codes on BI-NSO channels can then be stated as
follows.
 \begin{eqnarray}
    P^{(l)}(x)&=&P^{(0)}(x)\otimes
        \lambda\left(Q^{(l-1)}(x)\right)\label{eq:GDE-P}\\
    Q^{(l-1)}(x)&=&
         \Gamma^{-1}\left(\rho\left(\Gamma\left(\frac{P^{(l-1)}(0)+P^{(l-1)}(1)}{2}\right)\right)\right.\nonumber\\
     &+& \left.(-1)^x\rho\left(\Gamma\left(\frac{P^{(l-1)}(0)-P^{(l-1)}(1)}{2}\right)\right)
        \right).\label{eq:GDE-Q}
    \end{eqnarray}
The proportion of incorrect variable-to-check messages in the
$l$-th iteration can be computed by
\begin{eqnarray}
p_{e,linear}^{(l)}:=\int_{m=-\infty}^{0}\left(\frac{P^{(l)}(0)+P^{(l)}(1)}{2}\right)(dm).\nonumber
\end{eqnarray}
By iteratively computing $\left(P^{(l)}(0), P^{(l)}(1)\right)$ and checking
whether $p_{e,linear}^{(l)}$ converges to zero, we can determine whether the
channel of interest is asymptotically decodable when a sufficiently long linear
LDPC code is applied. Further discussion on the generalized DE can be found in
\cite{WangKulkarniPoor03}.

Let $\langle\cdot\rangle$ denote the average operator such that $\langle
f\rangle:=\frac{f(0)+f(1)}{2}$. One can easily show that the DE corresponding
to \Figure{\ref{fig:coset-code-arg}(b)} becomes
    \begin{eqnarray}
    P_{coset}^{(l)}&=&\left\langle P^{(0)}\right\rangle\otimes \lambda\left(Q_{coset}^{(l-1)}\right)\label{eq:density-evolution-P} \\
    Q_{coset}^{(l-1)}&=&\Gamma^{-1}\left(\rho\left(\Gamma\left(P_{coset}^{(l-1)}\right)\right)\right),\label{eq:density-evolution-Q}
    \end{eqnarray}
and the proportion of incorrect messages is
\begin{eqnarray}
p_{e,coset}^{(l)}:=\int_{m=-\infty}^{0}
P_{coset}^{(l)}(dm).\nonumber
\end{eqnarray}
One can check whether $p_{e,coset}^{(l)}$ converges to zero to determine
whether the channel of interest is asymptotically decodable with a sufficiently
long coset code ensemble. Further discussion on the coset-code-based approach
can be found in \cite{KavcicMaMitzanmacher03}.

\itwsection{Typicality of Linear LDPC Codes}

It was conjectured in \cite{HouSiegelMilisteinPfister03} that the scheme in
\Figure{\ref{fig:coset-code-arg}(c)} should have the same/similar
codeword-averaged  performance as those illustrated by
\Figures{\ref{fig:coset-code-arg}(a) {\rm and}~\ref{fig:coset-code-arg}(b)}. To
be more precise, the question is whether for the same channel model (namely,
for the same initial distribution pair $\left(P^{(0)}(0),P^{(0)}(1)\right)$)
 we are able to show
\begin{eqnarray}
\lim_{l\rightarrow\infty}p_{e,linear}^{(l)}=0\stackrel{?}{\Longleftrightarrow}
\lim_{l\rightarrow\infty}p_{e,coset}^{(l)}=0.\nonumber
\end{eqnarray}

This paper is devoted to the above question. We can answer immediately that the
performance of the linear code ensemble is very unlikely to be identical to
that of the coset code ensemble. However, when the minimum check node degree
$d_{c,min}:=\{k\in\NN:\rho_k>0\}$ is relatively large, we can prove that their
performance discrepancy is theoretically indistinguishable. In practice, the
discrepancy of decodable thresholds of the linear and the coset code ensemble
is within $0.05\%$ for a moderate $d_{c,min}\geq 6$.

\begin{figure}
{\hspace{-.3cm}\includegraphics[width=9.4cm,
keepaspectratio=true]{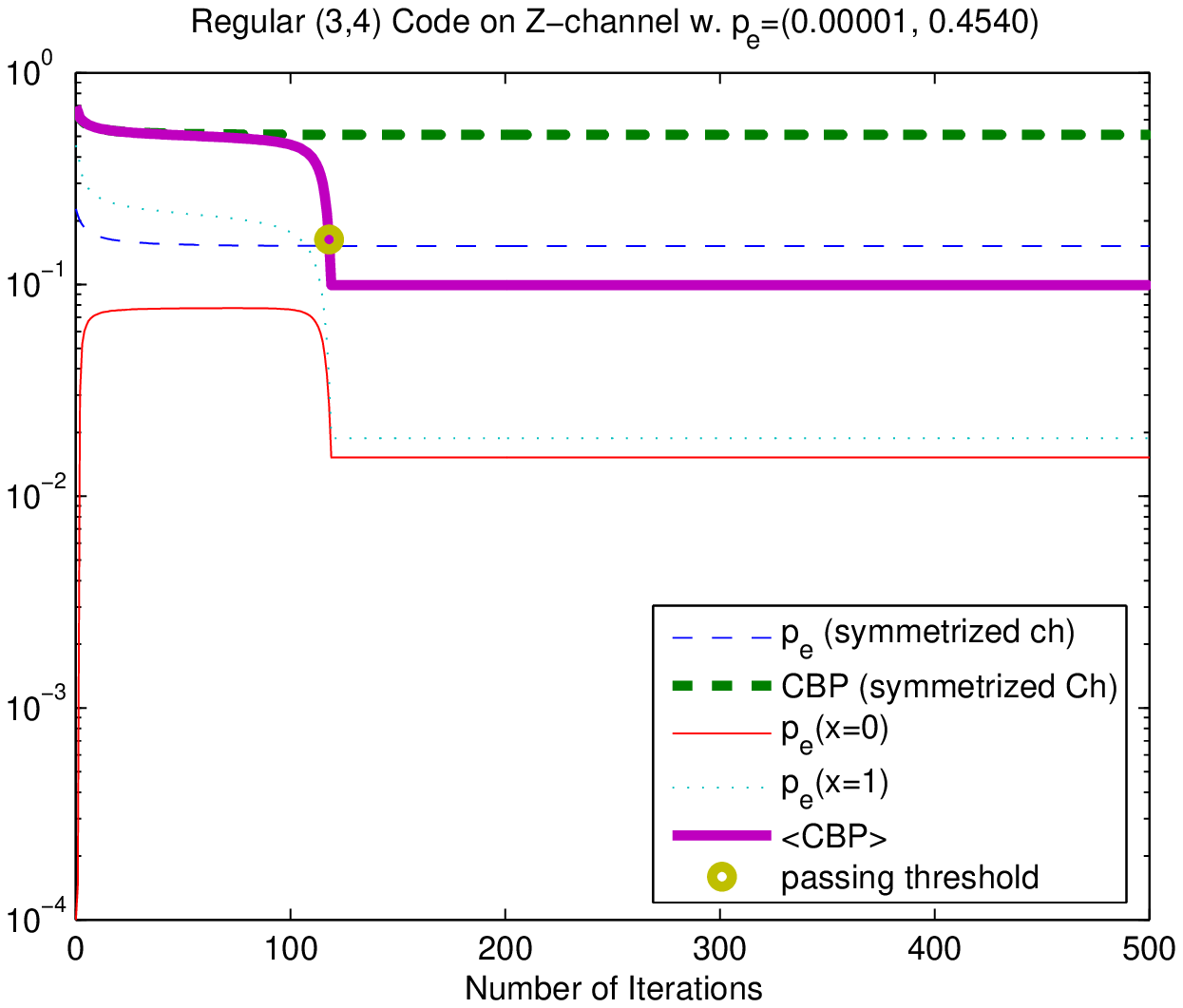}} \itwfigurecaption{Density evolution for
z-channels with the linear code ensemble and the coset code
ensemble.\label{fig:dens-evo-path}}
\end{figure}

It is clear from (\ref{eq:GDE-P}) that for linear codes, the
variable node iteration involves convolution of several densities
having the same $x$ value. The difference between $Q^{(l-1)}(0)$
and $Q^{(l-1)}(1)$ is thus amplified after each variable node
iteration. It is very unlikely that the decodable threshold of
linear codes (obtained from (\ref{eq:GDE-P}) and (\ref{eq:GDE-Q}))
and the decodable threshold of coset codes (obtained from
(\ref{eq:density-evolution-P}) and (\ref{eq:density-evolution-Q}))
will be analytically identical after the amplification during the
variable node iterations. \Figure{\ref{fig:dens-evo-path}}
demonstrates the traces of the evolved densities for the regular
(3,4) code on z-channels.\footnote{A z-channel is a
binary-input/binary-output channel such that only bit value 1 may
be contaminated with one-way crossover probability
$p_{1\rightarrow 0}$. Bit value 0 will always be received
perfectly.} With the one-way crossover probability being 0.4540,
the generalized DE for linear codes is able to converge within 179
iterations, while the coset code ensemble shows no convergence
within 500 iterations. This demonstrates the possible performance
discrepancy, though we do not have analytical results proving that
the latter will not converge after more iterations.
\Table{\ref{tab:typicality}} compares the decodable thresholds
such that the density evolution enters the stability region within
100 iterations. We notice that the larger $d_{c,min}$ is, the
smaller the discrepancy is. This phenomenon can be characterized
by the following theorem.
\begin{table*}
\centering \itwtablecaption{Threshold  comparison $p^*_{1\rightarrow 0}$ of
linear and coset LDPC codes on Z-channels\label{tab:typicality}}
\begin{tabular}{l|c|c|c|c}
\hline \hline
 $(\lambda,\rho)$       & $(x^2,x^3)$   & $(x^2,x^5)$   & $(x^2,0.5x^2+0.5x^3)$ & $(x^2,0.5x^4+0.5x^5)$\\
 \hline
Linear                  & 0.4540        & 0.2305        & 0.5888                & 0.2689\\
Coset                   & 0.4527        & 0.2304        & 0.5908                & 0.2690\\
\hline \hline
\end{tabular}
\end{table*}

\begin{theorem}\label{thm:typicality}
Consider BI-NSO channels and a fixed pair of finite-degree polynomials
$\lambda$ and $\rho$. The shifted version of the check node polynomial is
denoted by $\rho_{\Delta}=x^{\Delta}\cdot\rho$ where $\Delta\in\NN$. Let
$P^{(l)}_{coset}$ denote the evolved density from the coset code ensemble with
degrees $(\lambda,\rho_{\Delta})$ (obtained from (\ref{eq:density-evolution-P})
and (\ref{eq:density-evolution-Q})), and $\langle
P^{(l)}\rangle=\frac{1}{2}\sum_{x=0,1}P^{(l)}(x)$ denote the averaged density
from the linear code ensemble with degrees $(\lambda,\rho_{\Delta})$ (obtained
from (\ref{eq:GDE-P}) and (\ref{eq:GDE-Q})). Then, for any $l_0\in\NN$,
$\lim_{\Delta\rightarrow\infty}\langle P^{(l)}\rangle\stackrel{\mathcal
D}{=}P^{(l)}_{coset}$ in distribution for all $l\leq l_0$, with the convergence
rate being $\bigorder\left(\const^{\Delta}\right)$ for some $\const<1$.
\end{theorem}
\begin{corollary} [The Typicality on Z-Channels]\label{cor:typicality-z} Define
\begin{eqnarray}
p^*_{1\rightarrow 0, linear}&:=&\sup\left\{p_{1\rightarrow 0}>0:
\lim_{l\rightarrow\infty}p_{e,linear}^{(l)}=0\right\}\nonumber\\
\mbox{and~~}p^*_{1\rightarrow 0, coset}&:=&\sup\left\{p_{1\rightarrow 0}>0:
\lim_{l\rightarrow\infty}p_{e,coset}^{(l)}=0\right\}.\nonumber
\end{eqnarray}
For any $\epsilon>0$, there exists a $\Delta\in\NN$ such that
\begin{eqnarray}
\left|p^*_{1\rightarrow 0, linear}-p^*_{1\rightarrow 0,
coset}\right|<\epsilon.\nonumber
\end{eqnarray}
Namely, the asymptotic decodable thresholds of the linear and the coset code
ensemble are arbitrarily close when the minimum check node degree  $d_{c,min}$
is sufficiently large.
\end{corollary}
Similar corollaries can be constructed for other channel models
with different types of noise parameters. For example, the
$\sigma^*$ in the binary-input additive white Gaussian channel,
the $\lambda^*$ in the binary-input Laplace channel, etc. The
proof of \Corollary{\ref{cor:typicality-z}} is in
\Append{\ref{app:proof-typicality-z}}.

\begin{thmproof}{Proof of \Theorem{\ref{thm:typicality}}:}
Since the functionals in (\ref{eq:GDE-P}) and (\ref{eq:GDE-Q}) are
continuous with respect to convergence in distribution, we only
need to show that $\forall l\in\NN$,
\begin{eqnarray}
&&\lim_{\Delta\rightarrow\infty}Q^{(l-1)}(0)\stackrel{\mathcal D}{=}\lim_{\Delta\rightarrow\infty}Q^{(l-1)}(1)\nonumber\\
&&\stackrel{\mathcal
D}{=}~\Gamma^{-1}\left(\rho\left(\Gamma\left(\frac{P^{(l-1)}(0)+P^{(l-1)}(1)}{2}\right)\right)\right)\nonumber\\
&&=~\frac{Q^{(l-1)}(0)+Q^{(l-1)}(1)}{2}
,\label{eq:typicality-temp-1}
\end{eqnarray}
where $\stackrel{\mathcal D}{=}$ denotes convergence in
distribution. Then by inductively applying this weak convergence
argument, for any bounded $l_0$,
$\lim_{\Delta\rightarrow\infty}\langle
P^{(l)}\rangle\stackrel{\mathcal D}{=}P^{(l)}_{coset}$ in
distribution for all $l\leq l_0$. Without loss of
generality,\footnote{We also need to assume that $\forall x,
P^{(l-1)}(x)(m=0)=0$ so that
$\ln\coth\left|\frac{m}{2}\right|\in\RR^+$ almost surely. This
assumption can be relaxed by separately considering the event that
$m_{in,i}=0$ for some $i\in\{1,\cdots,d_c-1\}$.} we may assume
$\rho_{\Delta}=x^\Delta$ and prove the weak convergence of
distributions on the domain
\begin{eqnarray}
\gamma(m)&:=&\left(1_{\{m\leq0\}},\ln\coth\left|\frac{m}{2}\right|\right)=(\gamma_1,\gamma_2)\in\GF(2)\times
\RR^+,\nonumber
\end{eqnarray}
on which the check node iteration becomes
\begin{eqnarray}
\gamma_{out,\Delta}=\gamma_{in,1}+\gamma_{in,2}+\cdots +
\gamma_{in,\Delta}.\nonumber
\end{eqnarray}
Let $P'_0$ denote the density of $\gamma_{in}(m)$ given that the
distribution of $m$ is $P^{(l-1)}(0)$ and let $P'_1$ similarly
correspond to $P^{(l-1)}(1)$. Similarly let $Q'_{0,\Delta}$ and
$Q'_{1,\Delta}$ denote the output distributions on
$\gamma_{out,\Delta}$ when the check node degree is $\Delta+1$. It
is worth noting that any pair of $Q'_{0,\Delta}$ and
$Q'_{1,\Delta}$ can be mapped bijectively back to the LLR
distributions $Q^{(l-1)}(0)$ and $Q^{(l-1)}(1)$.

 Let
$\Phi_{P'}(k,r):=\EE_{P'}\left\{(-1)^{k\gamma_1}e^{ir\gamma_2}\right\},
\forall k\in\NN, r\in\RR$, denote the Fourier transform of the
density $P'$. Proving (\ref{eq:typicality-temp-1}) is equivalent
to showing that
\begin{eqnarray}
\forall k\in\NN, r\in\RR,~
\lim_{\Delta\rightarrow\infty}\Phi_{Q'_{0,\Delta}}(k,{r})=
\lim_{\Delta\rightarrow\infty}\Phi_{Q'_{1,\Delta}}(k,{r}).\nonumber
\end{eqnarray}
However, to deal with the strictly growing average of the limit distribution on
the second component of $\gamma_{out,\Delta}$, we concentrate instead on the
distribution of the normalized output $\left(\gamma_{1,out,\Delta},
\frac{\gamma_{2,out,\Delta}}{\Delta}\right)$. We then need to prove that
\begin{eqnarray}
\forall k\in\NN, r\in\RR,~
\lim_{\Delta\rightarrow\infty}\Phi_{Q'_{0,\Delta}}(k,\frac{r}{\Delta})=
\lim_{\Delta\rightarrow\infty}\Phi_{Q'_{1,\Delta}}(k,\frac{r}{\Delta}).\nonumber
\end{eqnarray}
We first note that $Q'_{0,\Delta}$ is the averaged distribution of
$\gamma_{out,\Delta}$ when the inputs $\gamma_{in,i}$ are governed
by $P^{(l-1)}(x_i)$ with $\sum_{i=1}^\Delta x_i=0$. Similarly
$Q'_{1,\Delta}$ is the averaged distribution of
$\gamma_{out,\Delta}$ when the inputs $\gamma_{in,i}$ are governed
by $P^{(l-1)}(x_i)$ with $\sum_{i=1}^\Delta x_i=1$. From the above
observation, we can derive the following iterative equations:
$\forall\Delta\in\NN$,
\begin{eqnarray}
\Phi_{Q'_{0,\Delta}}(k,\frac{r}{\Delta})&=&\frac{1}{2}\sum_{x=0,1}\Phi_{Q'_{x,\Delta-1}}(k,\frac{r}{\Delta})\Phi_{P'_x}(k,\frac{r}{\Delta})\nonumber\\
\Phi_{Q'_{1,\Delta}}(k,\frac{r}{\Delta})&=&\frac{1}{2}\sum_{x=0,1}\Phi_{Q'_{x,\Delta-1}}(k,\frac{r}{\Delta})\Phi_{P'_{\bar{x}}}(k,\frac{r}{\Delta}).\nonumber
\end{eqnarray}
 By induction, the difference thus becomes
\begin{eqnarray}
&&\Phi_{Q'_{0,\Delta}}(k,\frac{r}{\Delta})-\Phi_{Q'_{1,\Delta}}(k,\frac{r}{\Delta})\nonumber\\
&&=\left(\Phi_{Q'_{0,\Delta-1}}(k,\frac{r}{\Delta})-\Phi_{Q'_{1,\Delta-1}}(k,\frac{r}{\Delta})\right)\nonumber\\
&&~~~~\cdot        \left(\frac{\Phi_{P'_0}(k,\frac{r}{\Delta})-\Phi_{P'_1}(k,\frac{r}{\Delta})}{2}\right)\nonumber\\
&&=2\left(\frac{\Phi_{P'_0}(k,\frac{r}{\Delta})-\Phi_{P'_1}(k,\frac{r}{\Delta})}{2}\right)^\Delta.\label{eq:typicality-temp2}
\end{eqnarray}
By Taylor's expansion and the channel decomposition argument
in~\cite{WangKulkarniPoor04}, we can show that for all $k\in\NN$,
$r\in\RR$, and for all possible $P'_0$ and $P'_1$, the quantity in
(\ref{eq:typicality-temp2}) converges to zero with convergence
rate $\bigorder\left(\const^\Delta\right)$ for some $\const<1$. A
detailed derivation of the convergence rate is given in
\Append{\ref{app:conv-rate}}. Since the limit of the right-hand
side of (\ref{eq:typicality-temp2}) is zero, the proof of weak
convergence is complete. The exponentially fast convergence rate
$\bigorder\left(\const^\Delta\right)$ also justifies the fact that
even for moderate $d_{c,min}$ (e.g.\ $d_{c,min}\geq 6$), the
performances of linear and coset LDPC codes are very close.
\end{thmproof}

{\it Remark 1:} Consider any non-perfect message distribution, namely, $\exists
    x_0\in\{0,1\}$ such that $P^{(l-1)}(x_0)\neq \delta_{\infty}$, where $\delta_{m_0}$ is the Dirac delta measure
centered on $m_0$.    A persistent reader may notice that $\forall
x, \lim_{\Delta\rightarrow\infty}Q^{(l-1)}(x)\stackrel{\mathcal
D}{=}\delta_0$. That is, as $\Delta$ becomes large, all
information is erased after passing a check node of degree
$\Delta$. If this convergence (erasure effect) occurs earlier than
the convergence of $Q^{(l-1)}(0)$ and $Q^{(l-1)}(1)$, the
performances of linear and coset LDPC codes are ``close" only when
the corresponding codes are ``useless."\footnote{To be more
precise, they correspond to extremely high-rate codes and the
information is erased after every check node iteration.} To
quantify the convergence rate, we consider again the distributions
on $\gamma$ and their Fourier transforms. For the average of the
output distributions $\left\langle Q^{(l-1)}\right\rangle$, we
have
\begin{eqnarray}
&&\frac{\Phi_{Q'_{0,\Delta}}(k,\frac{r}{\Delta})+\Phi_{Q'_{1,\Delta}}(k,\frac{r}{\Delta})}{2}\nonumber\\
&&=\left(\frac{\Phi_{Q'_{0,\Delta-1}}(k,\frac{r}{\Delta})+\Phi_{Q'_{1,\Delta-1}}(k,\frac{r}{\Delta})}{2}\right)\nonumber\\
&&~~~~\cdot
        \left(\frac{\Phi_{P'_0}(k,\frac{r}{\Delta})+\Phi_{P'_1}(k,\frac{r}{\Delta})}{2}\right)\nonumber\\
        &&=\left(\frac{\Phi_{P'_0}(k,\frac{r}{\Delta})+\Phi_{P'_1}(k,\frac{r}{\Delta})}{2}\right)^\Delta.\label{eq:typicality-temp3}
\end{eqnarray}
By Taylor's expansion and the channel decomposition argument, one
can show that the limit of (\ref{eq:typicality-temp3}) exists and
the convergence rate is $\bigorder(\Delta^{-1})$. (A detailed
derivation is included in \Append{\ref{app:conv-rate}}.) This
convergence rate is much slower than the exponential rate
$\bigorder\left(\const^\Delta\right)$ in the proof of
\Theorem{\ref{thm:typicality}}. Therefore, we do not need to worry
about the case in which the required $\Delta$ for the convergence
of $Q^{(l-1)}(0)$ and $Q^{(l-1)}(1)$ is excessively large so that
$\forall x\in\GF(2), Q^{(l-1)}(x)\stackrel{\mathcal
D}{\approx}\delta_0$.

{\it Remark 2:} The intuition behind \Theorem{\ref{thm:typicality}} is that
when the minimum $d_c$ is sufficiently large, the parity check constraint
becomes relatively less stringent. Thus we can approximate the density of the
outgoing messages for linear codes by assuming all bits
$\{x_i\}_{i\in\{1,\cdots, d_c-2\}}$ involved in that particular parity check
equation are ``independently" distributed in $\{0,1\}$ rather than satisfying
$\sum x_i=x$, which leads to the formula for the coset code ensemble.
 On the other hand, extremely large $d_c$ is
required for a check node iteration to completely destroy all
information coming from the previous iteration. This explains the
difference between their convergence rates:
$\bigorder\left(\const^\Delta\right)$ versus
$\bigorder(\Delta^{-1})$.

\Figure{\ref{fig:weak-convergence}} illustrates the weak
convergence predicted by \Theorem{\ref{thm:typicality}} and
depicts the convergence rates of $Q^{(l-1)}(0)\longrightarrow
Q^{(l-1)}(1)$ and
$\frac{Q^{(l-1)}(0)+Q^{(l-1)}(1)}{2}\rightarrow\delta_0$.

\begin{figure*}
{\hspace{-.3cm}\includegraphics[width=9.6cm,
keepaspectratio=true]{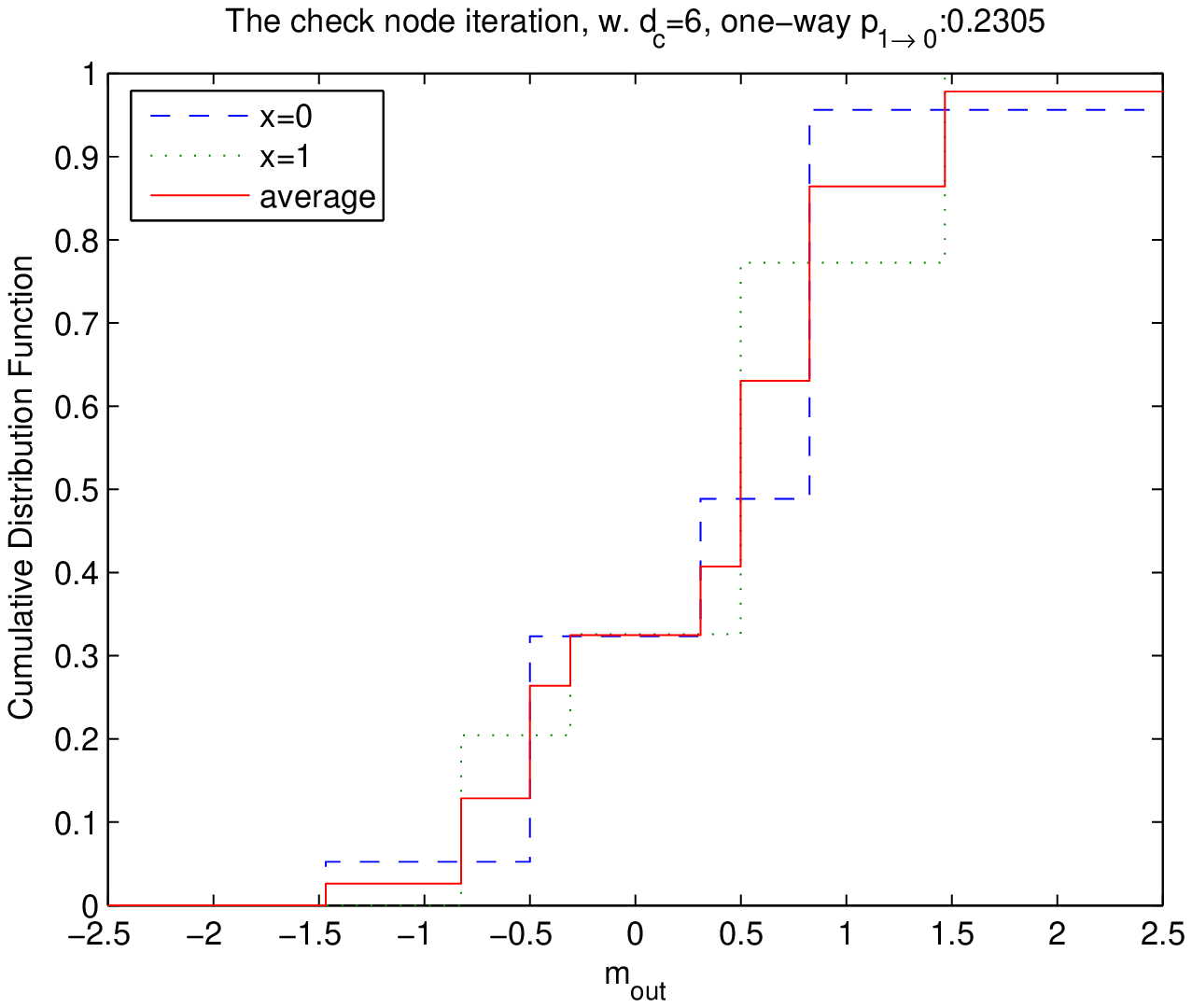}\hspace{-.7cm}\includegraphics[width=9.6cm,
keepaspectratio=true]{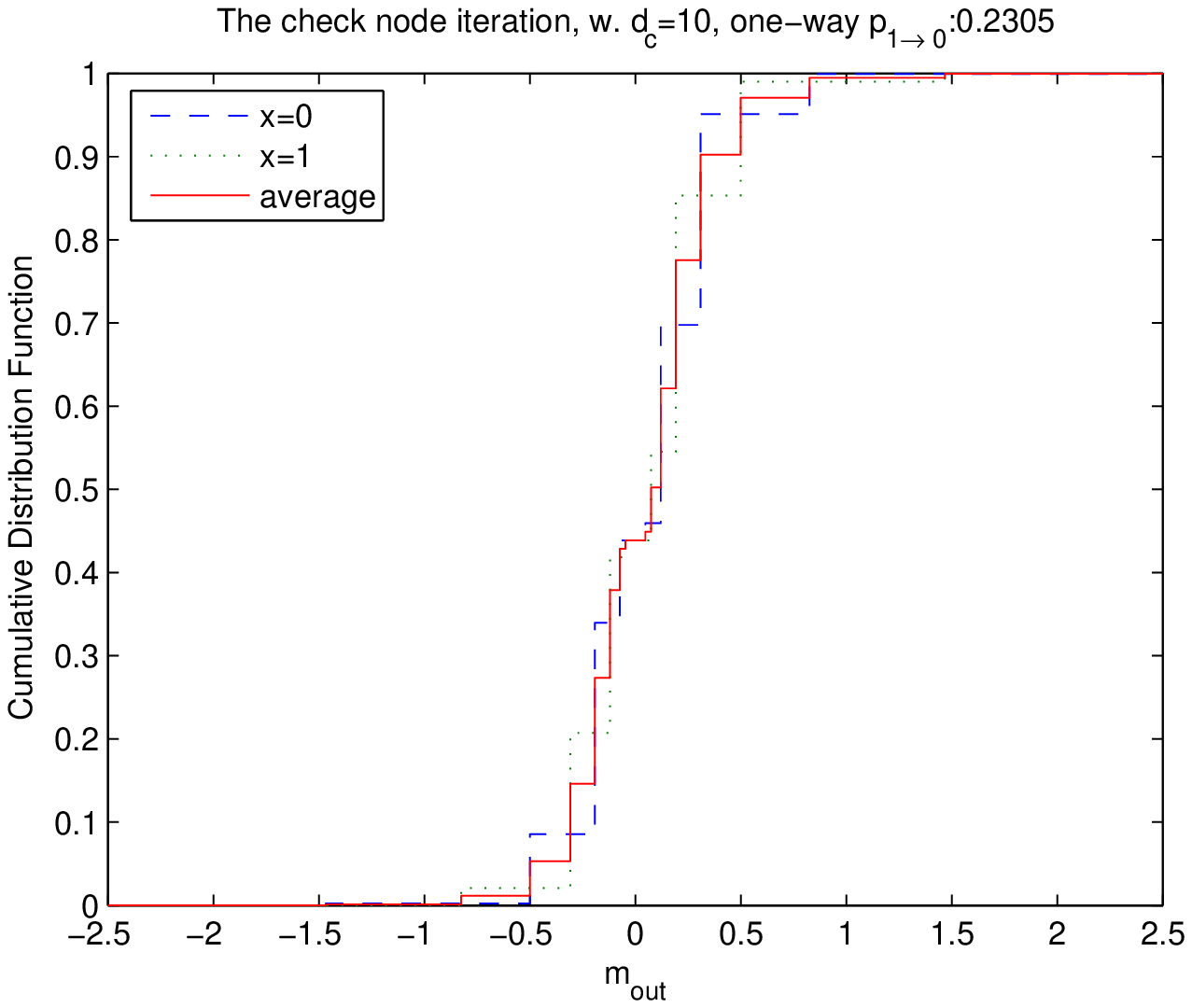}}
\itwfigurecaption{Illustration of the weak convergence of
$Q^{(l-1)}(0)$ and $Q^{(l-1)}(1)$. One can see that the
convergence of $Q^{(l-1)}(0)$ and $Q^{(l-1)}(1)$ is faster than
the convergence of $\frac{Q^{(l-1)}(0)+Q^{(l-1)}(1)}{2}$ and
$\delta_0$.\label{fig:weak-convergence}}
\end{figure*}

\itwsection{Conclusions\label{sec:conclusions}}

The typicality of the linear LDPC code ensemble has been proven by the weak
convergence (w.r.t.\ $d_c$) of the evolved densities in our codeword-averaged
density evolution. Namely, when the check node degree is sufficiently large
(e.g.\ $d_c\geq 6$), the performance of the linear LDPC code ensemble is very
close to (e.g.\ within $0.05\%$) the performance of the LDPC coset code
ensemble. This result can be viewed as a complementing theorem of the
concentration theorem in [{\it Corollary~2.2}
of~\cite{KavcicMaMitzanmacher03}], where a constructive method of finding a
typical coset-defining syndrome $\mathbf{s}$ is not specified.\footnote{Our
result shows the typicality of the all-zero $\mathbf s$ when $d_c$ is
sufficiently large. The results in \cite{KavcicMaMitzanmacher03}, on the other
hand, prove that a very large proportion of $\mathbf s$ is typical but $\mathbf
0$ may or may not be one of them.}

Besides the theoretical importance, we are then on a solid basis to
interchangeably use the linear LDPC codes and the LDPC coset codes when the
check node degree is of moderate size. For instance, from the implementation
point of view, the hardware uniformity of linear codes makes them a superior
choice compared to any other coset code. We can then use fast density evolution
\cite{JinRichardson04} plus the coset code ensemble to optimize the degree
distribution for the linear LDPC codes. Or instead of simulating the
codeword-averaged performance of linear LDPC codes, we can simulate the error
probability of the all-zero codeword in the coset code ensemble, in which the
efficient LDPC encoder \cite{RichardsonUrbanke01b} is not necessary.

\appendices

\itwsection{Proof of
\Corollary{\ref{cor:typicality-z}}\label{app:proof-typicality-z}} We prove one
direction that
\begin{eqnarray}
p^*_{1\rightarrow 0, linear}&>&p^*_{1\rightarrow 0, coset}-\epsilon.\nonumber
\end{eqnarray}
The other direction that $p^*_{1\rightarrow 0,
coset}>p^*_{1\rightarrow 0, linear}-\epsilon$ can be easily
obtained by symmetry. One prerequisite of the following proof is
that both the linear code and the coset code have the same
stability region~\cite{WangKulkarniPoor03}.

By definition, for any $\epsilon>0$, we can find a sufficiently large
$l_0<\infty$ such that for the one-way crossover probability $p_{1\rightarrow
0}:=p^*_{1\rightarrow 0, coset}-\epsilon$, $P^{(l_0)}_{coset}$ is in the
interior of the stability region. We first note that the stability region
depends only on the Bhattacharyya noise parameter~\cite{RichardsonUrbanke01a},
which is a continuous function with respect to convergence in distribution.
Therefore, by \Theorem{\ref{thm:typicality}}, there exists a $\Delta\in\NN$
such that $\left\langle P^{(l_0)}\right\rangle$ is also in the stability
region. By the definition of the stability region, we have
$\lim_{l\rightarrow\infty}p_{e,linear}^{(l)}=0$, which implies
$p^*_{1\rightarrow 0, linear}\geq p_{1\rightarrow 0}$. The proof is thus
complete.

\itwsection{The Convergence Rates of (\ref{eq:typicality-temp2})
and (\ref{eq:typicality-temp3})\label{app:conv-rate}} For
(\ref{eq:typicality-temp2}), we will consider the cases $k=0$ and
$k=1$ separately. By the binary asymmetric channel (BASC)
decomposition argument, namely, all binary-input non-symmetric
channels can be decomposed as the probabilistic combination of
many BASCs, we can limit our attention to simple BASCs rather than
general BI-NSO channels. Suppose
$\left(P^{(l-1)}(0),P^{(l-1)}(1)\right)$ corresponds to a BASC
with crossover probabilities $\epsilon_0$ and $\epsilon_1$.
Without loss of generality, we may assume $\epsilon_0+\epsilon_1<
1$ because of the previous assumption that $\forall x\in\GF(2),
P^{(l-1)}(x)(m=0)=0$. We then have
\begin{eqnarray}
\Phi_{P'_0}(k,\frac{r}{\Delta})=(1-\epsilon_0)e^{i\frac{r}{\Delta}\ln\frac{1-\epsilon_0+\epsilon_1}{1-\epsilon_0-\epsilon_1}}
+(-1)^k\epsilon_0e^{i\frac{r}{\Delta}\ln\frac{1+\epsilon_0-\epsilon_1}{1-\epsilon_0-\epsilon_1}}\nonumber\\
\Phi_{P'_1}(k,\frac{r}{\Delta})=(1-\epsilon_1)e^{i\frac{r}{\Delta}\ln\frac{1+\epsilon_0-\epsilon_1}{1-\epsilon_0-\epsilon_1}}
+(-1)^k\epsilon_1e^{i\frac{r}{\Delta}\ln\frac{1-\epsilon_0+\epsilon_1}{1-\epsilon_0-\epsilon_1}}.\nonumber
\end{eqnarray}

By Taylor's expansion, for $k=0$, (\ref{eq:typicality-temp2}) becomes
\begin{eqnarray}
        2\left(\frac{\Phi_{P'_0}(0,\frac{r}{\Delta})-\Phi_{P'_1}(0,\frac{r}{\Delta})}{2}\right)^\Delta \hspace{4.5cm}\nonumber\\
=
2\left(i\left(\frac{1-\epsilon_0-\epsilon_1}{2}\right)\left(\frac{r}{\Delta}\right)\ln\left(\frac{1-\epsilon_0+\epsilon_1}{1+\epsilon_0-\epsilon_1}\right)+\bigorder\left(\left(\frac{r}{\Delta}\right)^2\right)\right)^\Delta,\nonumber
\end{eqnarray}
which converges to zero with convergence rate
$\bigorder\left(\bigorder(\Delta)^{-\Delta}\right)$. For $k=1$, we have
\begin{eqnarray}
       2\left(\frac{\Phi_{P'_0}(1,\frac{r}{\Delta})-\Phi_{P'_1}(1,\frac{r}{\Delta})}{2}\right)^\Delta\hspace{3cm}\nonumber\\
=
2\left(\left(\epsilon_1-\epsilon_0\right)+\frac{i}{2}\left(\frac{r}{\Delta}\right)f_-(\epsilon_0,\epsilon_1)+\bigorder\left(\left(\frac{r}{\Delta}\right)^2\right)\right)^\Delta,\label{eq:diff-conv}
\end{eqnarray}
where
\begin{eqnarray}
f_-(\epsilon_0,\epsilon_1)\hspace{7.6cm}\nonumber\\
:=\left(
(1-\epsilon_0+\epsilon_1)\ln\frac{1-\epsilon_0+\epsilon_1}{1-\epsilon_0-\epsilon_1}-
(1+\epsilon_0-\epsilon_1)\ln\frac{1+\epsilon_0-\epsilon_1}{1-\epsilon_0-\epsilon_1}
\right).\nonumber
\end{eqnarray}
(\ref{eq:diff-conv}) converges to zero with convergence rate
$\bigorder(\const^\Delta)$, where $\const$ satisfies
$|\epsilon_1-\epsilon_0|<\const<1$. Since the convergence rate is
determined by the slower of the above two, we have proven that
(\ref{eq:typicality-temp2}) converges to zero with rate
$\bigorder(\const^\Delta)$ for some $\const<1$.

Consider (\ref{eq:typicality-temp3}). By the assumption that the
input is not perfect, we have $\max(\epsilon_0,\epsilon_1)>0$. For
$k=0$, by Taylor's expansion, we have
\begin{eqnarray}
        &&\left(\frac{\Phi_{P'_0}(0,\frac{r}{\Delta})+\Phi_{P'_1}(0,\frac{r}{\Delta})}{2}\right)^\Delta\nonumber\\
&&=~
\left(1+\frac{i}{2}\left(\frac{r}{\Delta}\right)f_+(\epsilon_0,\epsilon_1)+\bigorder\left(\left(\frac{r}{\Delta}\right)^2\right)\right)^\Delta,\label{eq:avg-conv}
\end{eqnarray}
where
\begin{eqnarray}
f_+(\epsilon_0,\epsilon_1)\hspace{7.6cm}\nonumber\\
:=\left(
(1-\epsilon_0+\epsilon_1)\ln\frac{1-\epsilon_0+\epsilon_1}{1-\epsilon_0-\epsilon_1}+
(1+\epsilon_0-\epsilon_1)\ln\frac{1+\epsilon_0-\epsilon_1}{1-\epsilon_0-\epsilon_1}
\right).\nonumber
\end{eqnarray}
The quantity in (\ref{eq:avg-conv}) converges to
\begin{eqnarray}
e^{i\left(\frac{r}{2}\right)f_+(\epsilon_1,\epsilon_2)}\nonumber
\end{eqnarray}
with rate $\bigorder\left(\Delta^{-1}\right)$. For $k=1$, we have
\begin{eqnarray}
        &&\left(\frac{\Phi_{P'_0}(1,\frac{r}{\Delta})+\Phi_{P'_1}(1,\frac{r}{\Delta})}{2}\right)^\Delta\nonumber\\
&&=~
\left(\left(1-\epsilon_0-\epsilon_1\right)\left(\frac{e^{i\frac{r}{\Delta}\ln\frac{1-\epsilon_0+\epsilon_1}{1-\epsilon_0-\epsilon_1}}
+e^{i\frac{r}{\Delta}\ln\frac{1+\epsilon_0-\epsilon_1}{1-\epsilon_0-\epsilon_1}}}{2}\right)\right)^\Delta,\nonumber
\end{eqnarray}
which converges to zero with rate
$\bigorder\left((1-\epsilon_0-\epsilon_1)^\Delta\right)$. Since
the overall convergence rate is the slower of the above two, we
have proven that the convergence rate is
$\bigorder\left(\Delta^{-1}\right)$.

\end{itwpaper}


%
%
~

~

\begin{itwreferences}

\bibitem{Gallager63}
R.~G. Gallager,
\newblock {\em Low-Density Parity-Check Codes},
\newblock Number~21 in Research Monograph Series. MIT Press, Cambridge, MA,
  1963.

\bibitem{McElieceMacKayCheng98}
R.~J. McEliece, D.~J.~C. Mackay, and J.~F. Cheng,
\newblock ``Turbo decoding as an instance of {Pearl's} {``Belief Propagation"}
  algorithm,''
\newblock {\em IEEE J. Select. Areas Commun.}, vol. 16, no. 2, pp. 140--152,
  Feb. 1998.

\bibitem{RichardsonUrbanke01a}
T.~J. Richardson and R.~L. Urbanke,
\newblock ``The capacity of low-density parity-check codes,''
\newblock {\em IEEE Trans. Inform. Theory}, vol. 47, no. 2, pp. 599--618, Feb.
  2001.

\bibitem{RichardsonShokrollahiUrbanke01}
T.~J. Richardson, M.~A. Shokrollahi, and R.~L. Urbanke,
\newblock ``Design of capacity-approaching irregular low-density parity-check
  codes,''
\newblock {\em IEEE Trans. Inform. Theory}, vol. 47, no. 2, pp. 619--637, Feb.
  2001.

\bibitem{WangKulkarniPoor03}
C.~C. Wang, S.~R. Kulkarni, and H.~V. Poor,
\newblock ``Density evolution for asymmetric memoryless channels,''
\newblock in {\em Proc. Int'l. Symp. Turbo Codes {\&} Related Topics}. Brest,
  France, 2003, pp. 121--124.

\bibitem{KavcicMaMitzanmacher03}
A.~Kav\v{c}i\'{c}, X.~Ma, and M.~Mitzenmacher,
\newblock ``Binary intersymbol interference channels: {Gallager} codes, density
  evolution and code performance bound,''
\newblock {\em IEEE Trans. Inform. Theory}, vol. 49, no. 7, pp. 1636--1652,
  July 2003.

\bibitem{HouSiegelMilisteinPfister03}
J.~Hou, P.~H. Siegel, L.~B. Milstein, and H.~D. Pfister,
\newblock ``Capacity-approaching bandwidth-efficient coded modulation schemes
  based on low-density parity-check codes,''
\newblock {\em IEEE Trans. Inform. Theory}, vol. 49, no. 9, pp. 2141--2155,
  Sept. 2003.

\bibitem{WangKulkarniPoor04}
C.~C. Wang, S.~R. Kulkarni, and H.~V. Poor,
\newblock ``On finite-dimensional bounds for {LDPC}-like codes with iterative
  decoding,''
\newblock in {\em Proc. Int'l Symp. Inform. Theory \& its Applications}. Parma,
  Italy, Oct. 2004.

\bibitem{JinRichardson04}
H.~Jin and T.~J. Richardson,
\newblock ``Fast density evolution,''
\newblock in {\em Proc. 38th Conf. Inform. Sciences and Systems}. Princeton,
  NJ, USA, 2004.

\bibitem{RichardsonUrbanke01b}
T.~J. Richardson and R.~L. Urbanke,
\newblock ``Efficient encoding of low-density parity-check codes,''
\newblock {\em IEEE Trans. Inform. Theory}, vol. 47, no. 2, pp. 638--656, Feb.
  2001.

\end{itwreferences}


\end{document}